\documentclass[twocolumn,showpacs,floatfix,prb]{revtex4}

\usepackage{amsmath,amssymb}
\usepackage{xspace}
\usepackage{graphicx}
\usepackage{dcolumn}
\usepackage{bm}

\input{MyMc.tex}

\newcommand{\ZeroPi}{{\ensuremath{0\text{-}\pi}}\xspace}

\begin{document}


\title{Static and dynamic properties 
of $0$, $\pi$, and \ZeroPi ferromagnetic tunnel Josephson Junctions}

\author{J. Pfeiffer}
\email{judith.pfeiffer@uni-tuebingen.de}
\author{M. Kemmler}
\author{D. Koelle}
\author{R. Kleiner}
\author{E. Goldobin}

\affiliation
{Physikalisches Institut-Experimentalphysik II and Center for Collective Quantum Phenomena, Universit\"at T\"ubingen, D-72076 T\"ubingen, Germany}
\author{M. Weides}

\affiliation
{Center of Nanoelectronic Systems for Information Technology (CNI), Research Center J\"ulich, D-52425 J\"ulich, Germany}
\author{A.~K.~Feofanov}
\author{J.~Lisenfeld}
\author{A. V. Ustinov} 

\affiliation{Physikalisches Institut III, Universit{\"a}t
 Erlangen-N{\"u}rnberg, D-91058 Erlangen, Germany
}%

\date{\today}

\begin{abstract}
  We present experimental studies of static and dynamic properties of $0$, $\pi$ and \ZeroPi superconductor-insulator-ferromagnet-superconductor (SIFS) Josephson junctions of small and intermediate length. In the underdamped limit these junctions exhibit a rich dynamical behavior such as resonant steps on the current-voltage characteristics. Varying the experimental conditions, zero field steps, Fiske steps and Shapiro steps are observed with a high resolution. A strong signature of the \ZeroPi Josephson junction is demonstrated by measuring the critical current as a function of two components ($B_x$, $B_y$) of an in-plane magnetic field. The experimental observation of a \textit{half-integer zero field step} in  \ZeroPi SIFS junctions is presented.

\end{abstract}


\maketitle

\section{Introduction}
\label{Sec:Intro}

The interplay between superconductivity and ferromagnetism has been studied during many decades \cite{Buzdin:2005:Review:SF}, but only during the last few years considerable results have been achieved regarding the experimental realization of $\pi$ and \ZeroPi Josephson junctions (JJs) using superconductor-ferromagnet (SF) multilayers. In such structures the Cooper pair wave function penetrates into the ferromagnet in the form of damped oscillations.\cite{fulde,larkin} If the thickness $d_F$ of the ferromagnetic barrier is of the order of half the oscillation period, the superconducting wave function changes its sign, \ie, shifts its phase by $\pi$ while crossing the ferromagnet. In SFS or SIFS JJs this leads to a negative critical current $I_c$ and the current-phase relation reads $I=I_c\sin(\phi)=|I_c|\sin(\phi+\pi)$ with $I_c<0$. Such a JJ is called ``$\pi$ JJ'', because it has $\phi=\pi$ in the ground state, \ie, when no bias current is applied. In this context, conventional JJs are called ``$0$ JJ'' because they have a current-phase relation of $I=I_c\sin\phi$ with $I_c>0$ and the ground state phase $\phi=0$. In experiment a single $0$ JJ and a single $\pi$ JJ are indistinguishable, as one can only determine $|I_c|$ because on current-voltage characteristics we have both $\pm I_c$, so that the sign of $I_c$ is invisible. During the last years several experiments showed a change in the sign of $I_c$ as a function of temperature\cite{Ryazanov:2001:SFS-PiJJ} $T$ or of the thickness of the ferromagnetic barrier\cite{Blum:2002:IcOscillations,Kontos:2002:SIFS-PiJJ,Oboznov:2006:SFS-Ic(dF),Weides:2006:SIFS-HiJcPiJJ} $d_F$. The signatures of $\pi$ JJs were also demonstrated by embedding SFS JJs into superconducting loops.\cite{Ryazanov:2002:SFS-PiArray,Guichard:2003:SFS:SQUID,Bauer:2004:SFS-SpontSuperCurrents} $\pi$ JJs are supposed to improve the performance of various classical and quantum electronic circuits, \eg, they can be used in RSFQ logics to self bias the circuit and to reduce the number of bias resistors\cite{Ortlepp:2006:RSFQ-0-pi} and inductances\cite{Ustinov:2003:RSFQ+pi-shifters} or to design environmentally decoupled ``quiet'' flux qubits.\cite{Ioffe:1999:sds-waveQubit,Yamashita:2005:pi-qubit:SFS+SIS} Depending on the target parameters, SFS or SIFS JJs are used. An advantage of SIFS JJs is that their resistance-area product $R \times A$ can be tuned over orders of magnitude by varying the thickness $d_I$ of the insulating barrier. A high $R \times A$ product facilitates voltage readout and allows to observe dynamic behavior of the junction due to low damping. Thus, these junctions can be used not only as phase batteries but also as active switching elements in superconducting electronic circuits. By contrast, SFS JJs have a small $R \times A$ product and are always in the overdamped limit. 

An intentionally made symmetric \ZeroPi SIFS JJ with two reference junctions was demonstrated recently.\cite{Weides:2006:SIFS-0-pi} Before that, \ZeroPi JJs were realized by utilizing d-wave superconductors\cite{Tsuei:Review,Smilde:ZigzagPRL,VanHarlingen:1995:Review,Hilgenkamp:zigzag:SF} or by chance using SFS JJs.\cite{DellaRocca:2005:0-pi-SFS:SF,Frolov:2006:SFS-0-pi} These SFS \ZeroPi JJs had a very small $R \times A$ product, an accidentally achieved $0$ to $\pi$ phase transition, and no reference junctions were available. The \ZeroPi SIFS JJ of Ref. \onlinecite{Weides:2006:SIFS-0-pi} consisted of $0$ and $\pi$ parts with equal critical current densities, as can be concluded from the measurements of the reference junctions. This \ZeroPi JJ had a high characteristic voltage $V_c=I_cR$ making direct transport measurements possible. The JJs could be driven into the underdamped regime by decreasing the temperature below 4.2\,K. Tuning the temperature, the same absolute critical current densities in both parts were achieved. In this symmetric case the ground state of the system consists of a spontaneously formed vortex of supercurrent circulating around the \ZeroPi boundary. This supercurrent corresponds to a local magnetic flux $|\Phi|\le\Phi_0/2$, where $\Phi_0=h/2e$ is the magnetic flux quantum. In the case of a very long \ZeroPi JJ the flux $\Phi=\pm\Phi_0/2$, and thus the localized magnetic field is called \emph{semifluxon}.\cite{Goldobin:SF-Shape} A \ZeroPi SIFS JJ is a promising system to study the physics of semifluxons. One can also use \ZeroPi-0 SIFS JJs to design semifluxon molecule qubits.\cite{Kato:1997:QuTunnel0pi0JJ,Goldobin:2005:MQC-2SFs} No restrictions in topography, low damping and good reproducibility make the phase modulated SIFS JJs good candidates for future logic elements.

In this paper we study dynamic and static properties of \ZeroPi SIFS JJs and of the corresponding reference $0$ and $\pi$ JJs. The junction parameters allow to investigate the overdamped as well as the underdamped regime by varying the temperature between $310\units{mK}$ and $4.2\units{K}$. Various resonant steps on the current-voltage characteristics are observed. Special attention is paid to the signatures of the $0$-$\pi$ boundary. 

\section{Experimental results}
\label{Sec:ExpRes}

\subsection{Samples}
\label{Sec:Samples}

\begin{figure}[!tb]
  \begin{center}
    \includegraphics{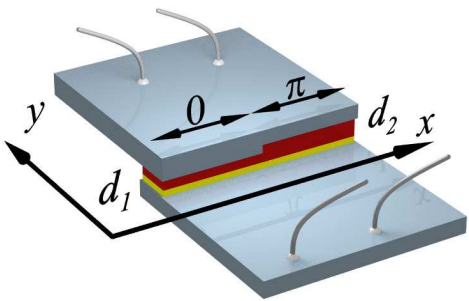}
  \end{center}
  \caption{(Color online) 
    Sketch of a SIFS \ZeroPi Josephson junction with a step-like change in the thickness $d_F$ (from $d_1$ to $d_2$) of the F-layer along the $x$ axis. Top and bottom electrode are shown by light blue (gray) color, the insulating barrier is yellow (light gray), the ferromagnetic barrier is red (dark gray).  
  }
  \label{Fig:Sketch}
\end{figure}

Our SIFS Josephson junctions are fabricated in overlap geometry using Nb/Al-Al$_2$O$_3$/Ni$_{60}$Cu$_{40}$/Nb technology.\cite{Weides:2007:JJ:TaylorBarrier,Weides:SIFS-Tech} Varying the thickness $d_F$ of the ferromagnetic barrier, Josephson junctions with $0$ and $\pi$ ground states are obtained.\cite{Weides:2006:SIFS-HiJcPiJJ} To fabricate \ZeroPi SIFS Josephson junctions, the ferromagnetic layer was selectively etched along one half of the junction. In this way one half of the junction has a F-layer thickness $d_F=d_1$ and, if taken separately, would be in a ground state with a phase drop of zero, while the other half of the junction has a F-layer thickness $d_F=d_2$ and, if taken separately, would be in a ground state with a phase drop of $\pi$.\cite{Weides:2006:SIFS-0-pi} A schematic drawing of such a \ZeroPi long JJ (LJJ) is shown in Fig.~\ref{Fig:Sketch}. The lengths of the $0$ and $\pi$ parts are equal with a lithographic accuracy of less than $1\units{\mu m}$. For each \ZeroPi JJ two reference junctions (one $0$ JJ with $d_F=d_1$ and one $\pi$ JJ with $d_F=d_2$), having the same length $L$ as the 0-$\pi$ JJ, are fabricated in the same run. We assume that the critical current densities $j_c^0$ and $j_c^{\pi}$ are the same in the reference and in the 0-$\pi$ JJs, so that the reference JJs can indeed be used to obtain information about the 0 and $\pi$ parts of a 0-$\pi$ JJ. 

In this paper we present experimental data for two sets of samples. Each set is situated on a separate chip (chip $\#1$ and chip $\#2$) and contains three JJs: a \ZeroPi JJ, a 0 JJ and a $\pi$ JJ. The thickness of the insulating barrier is thinner on chip $\#2$, resulting in higher critical current densities in comparison with chip $\#1$. The junction parameters are summarized in Tab.~\ref{Tab:Params}. The critical current densities of the reference junctions are obtained by measuring their $I$-$V$ characteristics (IVCs), while for the \ZeroPi JJs only the average value $j_c^{\ZeroPi}=(|j^{\pi}_c| + j^0_c)/2$ is quoted. $j^{\ZeroPi}_c$ is used to calculate the normalized lengths $l$ of the \ZeroPi JJs. Actually, this $j^{\ZeroPi}_c$ value is a simplified picture to facilitate the calculations. It is different from the measured critical current density in a \ZeroPi JJ, \ie, from $j^{\ZeroPi}_c$ at $B=0$, a situation when the critical currents in both halves are partially or totally canceling each other. While calculating $l$ the idle region corrections are taken into account.\cite{Wallraff:PhD,Monaco:1995:IdleReg:Dyn}

\begin{table*}
  \centering
  \begin{tabular}{l|c c c c c c c}
    \hline\hline    
    \textit{id} @ $T$ (K) 
    & \ensuremath{j_c} ($\units{A/cm^2}$) 
    & \ensuremath{l}   
    & \ensuremath{L} (\units{\mu m}) 
    & \ensuremath{W_j} (\units{\mu m}) 
    & \ensuremath{W_i} (\units{\mu m})\\ 
    \hline\hline
    $\#1$-$0$ @ 4.2 K     &2.1  &0.72 &330  &30 &50 \\
    $\#1$-$\pi$  @ 4.2 K    &1.5  &0.62 &330  &30 &50 \\
    $\#1$-\ZeroPi @ 4.2 K   &1.8  &0.67 &330  &30 &50 \\
    \hline
    $\#2$-$0$ @ 0.34 K    &13.4 &3.1  &500  &12.5 &10 \\
    $\#2$-$\pi$ @ 0.34 K    &4.5  &1.8  &500  &12.5 &10 \\
    $\#2$-\ZeroPi @ 0.34 K  &9.0  &2.5  &500  &12.5 &10 \\
    \hline
  \end{tabular}
  \caption{%
    Parameters of the investigated samples at measurement temperatures. $L$ is the length of the junction, $W_{j,i}$ is the width of the junction or the idle region in physical units. $l$ is the normalized length of the respective JJ. Notations are chosen according to Ref.~\onlinecite{Monaco:1995:IdleReg:Dyn}.
  }
  \label{Tab:Params}
\end{table*}

Note that early measurements using sample set $\#1$ were already published.\cite{Weides:2006:SIFS-0-pi} The measurements reported here were done at least eight months later. It turned out that some parameters, \eg, $j_c$ values, slightly changed with time presumable due to the clustering in the F-layer and degradation of the interfaces.

The measurements are carried out in a standard $^4$He- or $^3$He cryostat. The $^4$He cryostat can be evacuated so that temperatures from $4.2\units{K}$ down to $2.1\units{K}$ are reached. Using the $^3$He cryostat, temperatures between $1.9\units{K}$ and $310\units{mK}$ are accessible. In both measurement setups a cryoperm shield is placed around the sample to shield it from the earth magnetic field or stray fields.

\subsection{Static properties}
\label{Sec:StaticProp}

\begin{figure}[tb]
  \includegraphics[width=8.6cm]{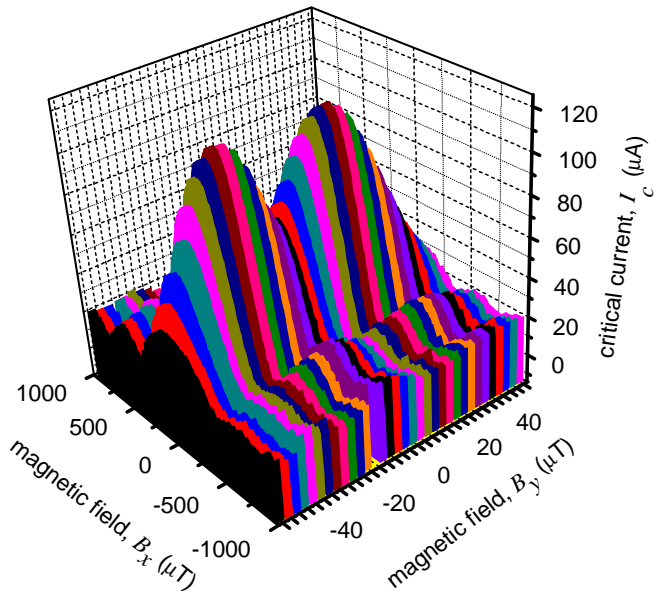}
  \caption{(Color online) 
    Experimentally measured $I_c^{\ZeroPi}(B_x,B_y)$ of the sample $\#1$-\ZeroPi at $T=4.2\units{K}$. For $B_y=0$, $I^{\ZeroPi}_c(B_x)$ shows a regular Fraunhofer pattern. For $B_x=0$, in $I^{\ZeroPi}_c(B_y)$ a minimum around $B_y\sim 0$ is visible which is a characteristic feature of \ZeroPi JJs. }
  \label{Fig:Ic(B)}
\end{figure}

We study the static properties of sample set $\#1$ by measuring the dependences of the critical current on magnetic field $I_c(B)$. The junctions are cooled down in the absence of magnetic field or bias current to provide a flux free state. Magnetic fields with both $x$ and $y$ components can be applied in the plane of the junctions, see Fig.~{\ref{Fig:Sketch}}. The $I^{0}_c(B_x,B_y)$ and $I^{\pi}_c(B_x,B_y)$ dependences of the $0$ and $\pi$ reference junctions have almost perfect Fraunhofer patterns (not shown) indicating a state without trapped flux. The maxima of the curves reveal a small offset from zero magnetic field probably due to a weak net magnetization of the ferromagnet. Applying only $B_x$, the field value at the first minimum of $I_c(B_x)$ dependence is $\sim 11$ times larger than the corresponding value on the $I_c(B_y)$ curve. This corresponds to the ratio between the length $L$ and the width $W_j$ of the Josephson junctions on chip $\#1$ ($L/W_j=11/1$).

Fig.~\ref{Fig:Ic(B)} shows the $I^{\ZeroPi}_c(B_x,B_y)$ dependence of the \ZeroPi junction of set $\#$1. Applying a magnetic field in $x$ direction results in an almost perfect Fraunhofer pattern, indicating that no parasitic flux is trapped in the junction or its electrodes. Applying magnetic field in $y$ direction a well pronounced minimum around zero field is visible. This minimum is a characteristic feature of a \ZeroPi Josephson junction.\cite{Wollman:1993:0-pi-JJ,Smilde:ZigzagPRL,Kirtley:IcH-PiLJJ} Thus, Fig.~\ref{Fig:Ic(B)} leaves no doubts that the observed behavior is due to the step in the ferromagnetic barrier and not due to some other reasons. The normalized junction length is calculated as $l=0.67$ at $T=4.2\units{K}$. In contrast to earlier investigations of this sample\cite{Weides:2006:SIFS-0-pi} the idle region corrections\cite{Wallraff:PhD,Monaco:1995:IdleReg:Dyn} are now taken into account. Although the junction is short in terms of $\lambda_J$ ($l<1$), $I_c$ in the central minimum on the $I_c^{\ZeroPi}(B_y)$ dependence does not reach zero as it should be for a short JJ\cite{Goldobin:Art-0-pi}, most probably because $j^0_c\neq j^{\pi}_c$ at $T=4.2\units{K}$. The asymmetry factor is estimated as $\Delta=j^{\pi}_c/j^0_c=0.71$, \ie, the critical current densities in the $0$ and $\pi$ parts differ by about 30\,$\%$. For such an asymmetry the ground state is fluxless \cite{Martin_neu,Bulaevski}.

\subsection{Dynamic properties}
\label{Sec:DynProp}

We study the dynamic properties by measuring current-voltage ($I$-$V$) characteristics (IVC). The IVC directly reveals the relationship between the average flux velocity $v$, which is proportional to the measured dc voltage $V$, and the driving force, which is proportional to the bias current $I$. To observe dynamics, the junction under examination has to be in the underdamped regime. This is typically the case for temperatures below $\sim 3\,{\rm K}$. At $T=4.2\units{K}$ none of our samples is underdamped. By decreasing the temperature all junctions become underdamped.

\subsubsection{Half-integer ZFSs in \ZeroPi junctions}
\label{Sec:HIZFS}

\begin{figure}[t]
  \includegraphics[width=7.6cm]{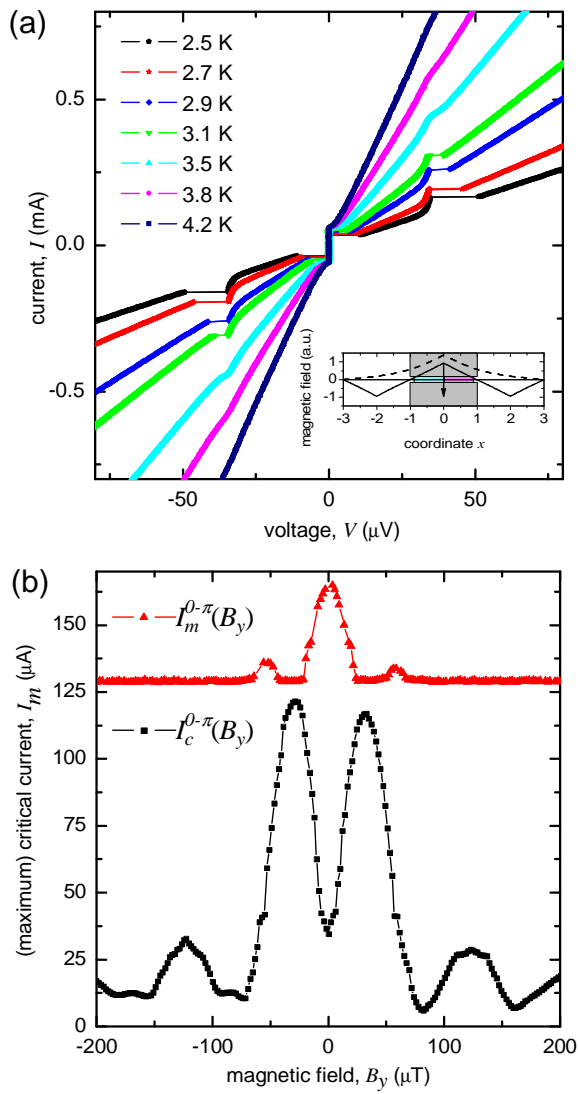}
  \caption{(Color online) 
    (a) Experimentally measured IVCs of JJ $\#1$-\ZeroPi at different temperatures. At $T<3.8\units{K}$ the half-integer zero field step is emerging. The inset shows the sketch of a \ZeroPi JJ with a fractional vortex which is permanently flipping. 
    (b) The dependence of the maximum current $I^{\ZeroPi}_m$ of the first half-integer ZFS at $V=35\units{\mu V}$ on magnetic field $B_y$,  measured at $T=2.5\units{K}$. The $I^{\ZeroPi}_c(B_y)$ dependence is shown for comparison.
  }
  \label{hizfs_mitIch}
\end{figure}

Fiske steps (FSs) appear on the IVCs of JJs in magnetic field when the parameter $l/\sqrt{\beta_c}$ is relatively small. Physically, a Fiske resonance is the synchronization between a moving Josephson vortex chain and a standing electromagnetic wave in the JJ. The resonance number $n$ determines the number of wave lengths of the standing wave. The asymptotic voltage of the $n$-th Fiske step is given by 
\begin{equation}
  V_n^\mathrm{FS}=n\frac{\Phi_0\bar{c}}{2 L},
\label{Eq:FS:Vn}
\end{equation}

where $\bar{c}$ is the Swihart velocity. By contrast, if IVC measurements are carried out in the absence of magnetic field or microwaves one can also observe steps on the IVC of a LJJ. These steps, naturally called zero field steps (ZFSs), have a voltage spacing twice larger than the one of Fiske steps.

The origin of ZFSs is different from Fiske steps. A ZFS with index $n$ corresponds to $n$ fluxons moving inside the JJ driven by the bias current. When one of the fluxons reaches the JJ boundary, it reflects back as an antifluxon, thus transferring a flux $\Phi=+\Phi_0-(-\Phi_0)=2\Phi_0$ through the boundary per $2L/v$ seconds --- the time needed for a fluxon to shuttle once back and forth along the JJ. Thus the voltage across the junction is given by
\begin{equation}
  V_n^{ZFS}=\frac{\Delta\Phi}{\delta t}=n\frac{2\Phi_0 v}{2L}=
  n\frac{\Phi_0 v}{L}
  . \label{Eq:ZFS:V}
\end{equation}

If, in this mode, one increases the bias current, the fluxons move faster and eventually almost reach the Swihart velocity ${\bar c}$. Thus, the voltage spacing between the ZFSs is $\Delta V^\mathrm{ZFS}=\Phi_0 {\bar c}/L$, \ie, twice larger than for Fiske steps.

These expectations are met in our experiments in the underdamped limit with the $0$ and $\pi$ junction of sample set $\#1$. The observed ZFSs have a voltage spacing which is exactly two times larger than the one of Fiske steps (not shown). 

In the case of the \ZeroPi junction the first ZFS emerges at exactly \textit{half} of the usual ZFS spacing, \ie, exactly at \textit{the same} voltage as the first Fiske step, according to the earlier predictions\cite{Stefanakis:ZFS/2,Lazarides:Ic(H):SF-Gen} and to experiments on \ZeroPi JJs of other types.\cite{Goldobin:Art-0-pi} Fig.~\ref{hizfs_mitIch} (a) shows IVCs of the JJ $\#1$-\ZeroPi measured at different $T$ and at $B=0$. At $35\units{\mu V}$, which is exactly the observed value of the Fiske step voltage spacing in this sample, a step is emerging for $T<3.8\units{K}$. The step appears due to the flipping of a fractional vortex.\cite{Goldobin:Art-0-pi} At zero bias current, for our JJ length and $j_c$ asymmetry, the ground state of the system is a flat phase state $\phi=0$ (fluxless). As soon as a uniform bias current is applied, a fractional flux (Josephson vortex) localized at the 0-$\pi$ boundary and carrying a bias current dependent flux appears.\cite{Goldobin:SF-ReArrange} Since the junction has a finite length, the interaction of this vortex with the boundary can be treated as the interaction of the vortex with an antivortex (image) situated outside the junction at the same distance from the edge. There are two such images, one behind the left and one behind the right edge of the JJ, see the inset of Fig.~\ref{hizfs_mitIch} (a). The bias current exerts a Lorenz force which tries to collide the vortex with one of the images. If the bias is large enough, both the vortex and the image vortex flip, changing polarity and exchanging one flux quantum, which, in fact, is passing through the JJ boundary. Then a similar process takes place between the fractional vortex (now of negative polarity) and another image, so one $\Phi_0$ passes through the other JJ boundary. Assuming that the maximum velocity of flux transfer is ${\bar c}$, we calculate that the asymptotic voltage is exactly equal to the voltage of the first Fiske step.

In order to show that the observed step is indeed a ZFS, the maximum current $I^{\ZeroPi}_m$ of this step versus magnetic field $B_y$ is measured, see Fig.~{\ref{hizfs_mitIch} (b). The $I_m^{\ZeroPi}(B_y)$ has a maximum in zero field and is decreasing with applied magnetic field --- which is the typical behavior of a ZFS. The background value of $I^{\ZeroPi}_m$ corresponds to the current on the McCumber branch at $V\approx35\units{\mu V}$ below which the step cannot be suppressed in principle. Fig.~{\ref{hizfs_mitIch} (b) additionally shows the $I_c^{\ZeroPi}(B_y)$ curve measured at the same temperature of $T=2.5\units{K}$ for comparison. The minimum around $B_y\approx0$, typical for 0-$\pi$ JJs, is visible. Note that the amplitude of $I_c(B_y)$ is higher than in Fig.~\ref{Fig:Ic(B)} as the temperature is lower. The next ZFS emerges at $\sim 100\units{\mu V}$ ($n=3/2$) and corresponds to one additional fluxon moving inside the \ZeroPi JJ. Comparable measurements are performed for sample set $\#2$ and show similar results (not shown).

Up to now half-integer ZFS have been observed only in artificial \ZeroPi junctions.\cite{Goldobin:Art-0-pi} To our knowledge, here we report on the first measurements of a half-integer ZFS in \ZeroPi SIFS junctions.

\subsubsection{Fiske steps}
\label{Sec:Fiske}

The junctions of sample set $\#2$ are prepared in a flux free state. By applying a magnetic field $B_y$ various Fiske steps were observed on the IVCs. Data presented below are taken at $T=340\units{mK}$ (underdamped regime) for sample $\#2$-$\pi$.

\begin{figure}[tb]
  \includegraphics[width=7.6cm]{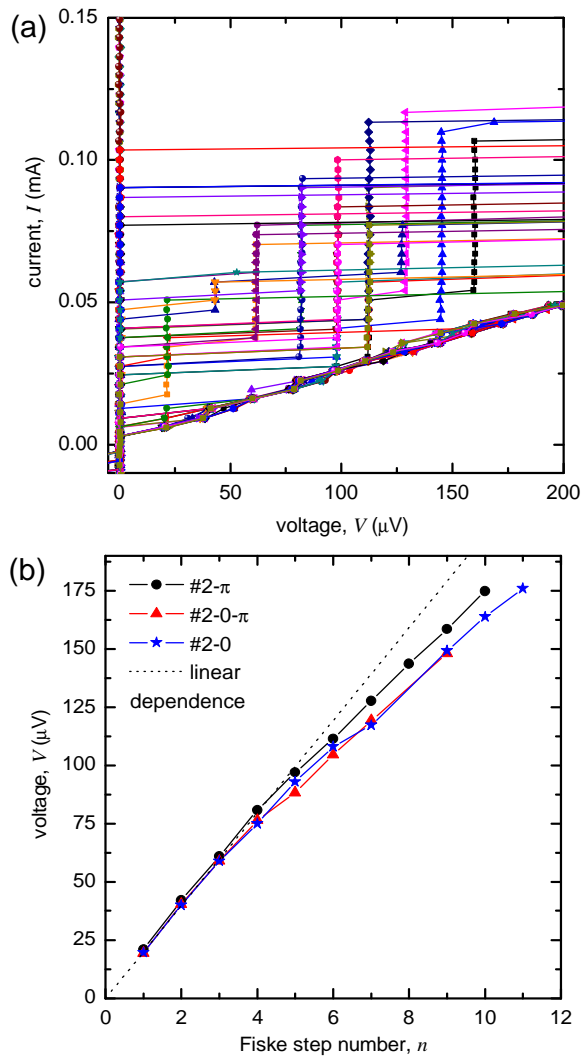}
  \caption{(Color online) 
    (a) Fiske steps on the IVCs of the junction $\#2$-$\pi$ at $T=340\units{mK}$. The magnetic field is varied between $B=-105\ldots105\units{\mu T}$. 
    (b) Asymptotic voltage of the $n$-th Fiske step is plotted versus $n$ for the sample set $\#2$. Taking into account only the first four Fiske steps $\Delta V=19.9\units{\mu V}$.
  }
  \label{fiske_m11_5_9_07fiske_number}
\end{figure}
 
The low voltage part of several IVCs of junction $\#2$-$\pi$ at different magnetic fields in the range $-105\ldots105\units{\mu T}$ is shown in Fig.~\ref{fiske_m11_5_9_07fiske_number} (a). Nine Fiske steps are observed with a high resolution. Similar Fiske step measurements were also carried out for the samples $\#2$-\ZeroPi and $\#2$-$0$ (not shown). To summarize these measurements, in Fig.~\ref{fiske_m11_5_9_07fiske_number} (b) we plot the voltage positions of Fiske steps versus step number for all three samples. The first four Fiske steps have an almost equidistant voltage spacing. For higher Fiske steps the voltage spacing between adjacent Fiske steps shrinks with increasing step number. Thus the dispersion relation of electromagnetic waves in the junctions is not linear. A possible explanation might be the layout of the junctions and the fabrication process: inhomogeneities in the junction may result in a decreasing voltage spacing of resonant steps on the IVC, compare with Ref. \onlinecite{Barbara:1996:IdleReg:FineZFS}. Local inhomogeneities are, \eg, non ideal junction boundaries, where the profile of the critical current does not sharply go to zero but is smeared out. This is the case if the junction is surrounded by a large idle region, which is true for our samples.

As visible in Fig.~\ref{fiske_m11_5_9_07fiske_number} (b), the Fiske step voltage spacing of the $\pi$ junction shows the smallest deviation from an equidistant voltage spacing, indicating a better homogeneity as compared to the $0$ and \ZeroPi junction. Note that the F-layer of the last two junctions (or parts of it) is etched during the fabrication process. The etching might cause additional inhomogeneities. This topic was already discussed in the literature as {\it material dispersion} \cite{Hermon,Lee}. The authors had shown that due to idle region effects or a frequency dependent magnetic penetration depth, summarized as material dispersion, the dispersion relation of electromagnetic waves in JJs is not linear.

Another explanation is that $j_c^0$ is more sensitive to variations in $d_F$ (roughness) than $j_c^{\pi}$ because $d_1$ and $d_2$ are chosen so that 
\begin{equation}
  \left| \fracp{I_c(d_F)}{d_F} \right|_{d_F=d_1} > 
  \left| \fracp{I_c(d_F)}{d_F} \right|_{d_F=d_2}
  , \label{Eq:Ic(d_F):DiffSlopes}
\end{equation}
see Fig.~1 of Ref.~\onlinecite{Weides:2006:SIFS-HiJcPiJJ}.

To calculate the capacitance of the junctions the voltage spacing of the first four Fiske steps is used. The capacitance is estimated as $C \approx 22.9\pm0.8\units{\mu F/cm^2}$ which also includes idle region effects. From geometrical considerations the contributions of the idle region ($C_i$) and  the \textit{naked} junction ($C_j$) to the capacitance $C=C_i+C_j$ can be estimated. The respective capacitance is calculated as $C_{i,j}=\epsilon_0\epsilon_{i,j}W_{i,j}L/d_I^{i,j}$, $L$ being the length and $W_{i,j}$ being the width of the junction or idle region, see Tab.~\ref{Tab:Params}. 

Taking $\epsilon_i\sim 30$, $\epsilon_j\sim10$ and $d_I^i=120\units{nm}$ and $d_I^j=0.4\units{nm}$, the idle region has a capacitance of $C_i=221.5\units{nF/cm^2}$. The Swihart velocity in the naked junction is calculated as $c_0=c\sqrt{d_I^j/\epsilon_j d_j^{`}}$ with $c$ being the vacuum speed of light and $\mu_0d_j^{`}$ being the inductance (per square) of the junction electrodes, resulting in $c_0=0.015c$. 

\begin{figure}[tb]
  \includegraphics[width=7.6cm]{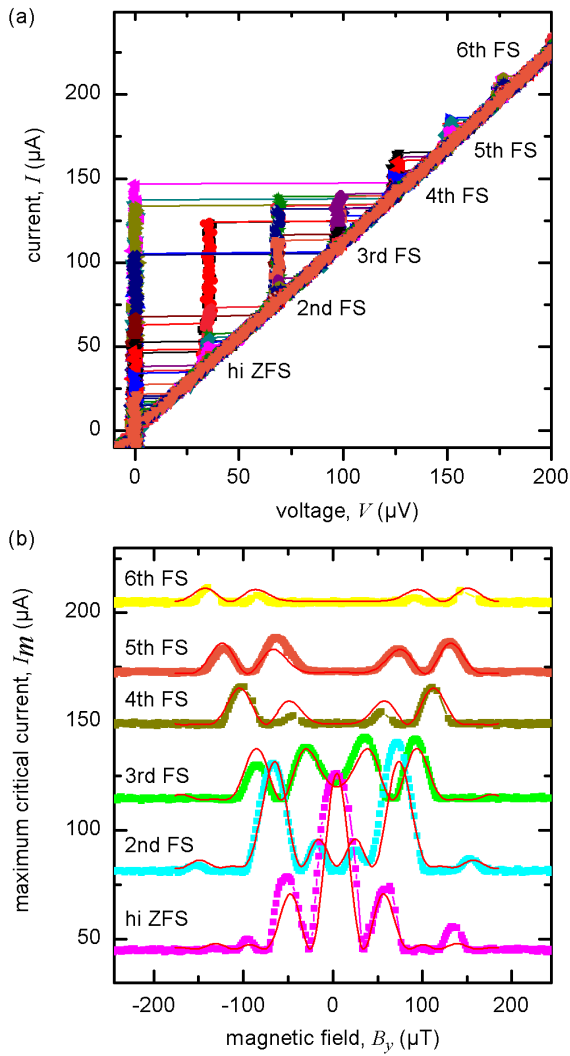}
  \caption{(Color online) 
    (a) Experimental data of Fiske steps on the IVCs of the junction $\#1$-\ZeroPi at $T=310\units{mK}$. The magnetic field is varied between $B=-200\ldots 200\units{\mu T}$. hi= half-integer.
    (b) Maximum current amplitude of the $n$th Fiske step versus applied magnetic field at $T=310\units{mK}$. Solid line: Analytical calculations performed according to Ref.\onlinecite{Nappi:2007:0-pi:Fiske} using the experimental parameters of sample $\#1$-\ZeroPi at $T=310\units{mK}$. }
  \label{fiske_PRL}
\end{figure}

Comparable Fiske step measurements are performed for sample set $\#1$. The low voltage part of several IVCs of sample $\#1$-\ZeroPi at $T=310\units{mK}$ is shown in Fig.~\ref{fiske_PRL} (a). The magnetic field is varied between $B=-200\ldots 200\units{\mu T}$. In Fig.~\ref{fiske_PRL} (b) the dependences of the maximum current of the Fiske steps on magnetic field are presented. Note that every odd FS is mixed with a half-integer ZFS, thus resulting in a finite step height even for $B=0$. Our experimental data reproduce theoretical predictions rather well\cite{Nappi:2007:0-pi:Fiske}. The authors used a perturbative scheme to calculate phase dynamics and the resulting resonances appearing on the IVC of JJs with an arbitrary number of \ZeroPi singularities. As realized in our samples they considered the flat phase regime where no spontaneous flux is occurring in the ground state of a JJ with finite length. Using the expressions number (27) and (28) from Ref. \onlinecite{Nappi:2007:0-pi:Fiske} with the respective parameters of our samples we are able to compare experiment and theory, see solid lines in Fig.~\ref{fiske_PRL} (b). The analytical calculations are performed in normalized units and converted to physical units by a fitting procedure: The $x$-axis is adjusted by superimposing the first minimum of the $I_m^{0-\pi}(B_y)$ curve of the half-integer ZFS in experiment and theory. This conversion scheme is used for all six measurements. The $y$-axes are adjusted separately using a similar procedure. We find a very good agreement between the analytical predictions and the experimental data. The spacing of the minima, the height of the maxima as well as the overall shape of the measurements are reproduced by theory\cite{Nappi:2007:0-pi:Fiske}.

As it was pointed out each odd step is a mixture of Fiske and half-integer ZFS. The amplitude $I_m(0)$ of the 3rd step is above the background level due to its ZFS contribution. For the 5th FS, $I_m(0)$ coincides with the background level as the ZFS contribution is vanishing.

\subsubsection{Shapiro steps}
\label{Sec:Shapiro}

\begin{figure}[t]
  \includegraphics[width=7.6cm]{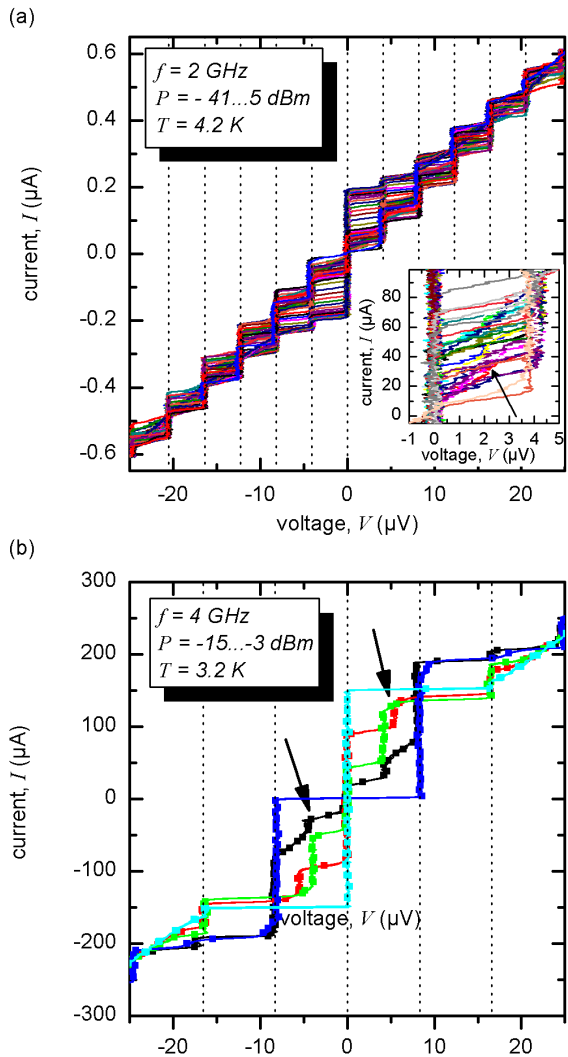}
  \caption{(Color online) 
    (a) Shapiro steps measured for junction $\#1$-$0$ at $T=4.2\units{K}$ (overdamped regime). Microwaves are applied with a frequency of $f=2\units{GHz}$, the power is varied between $P=-41\ldots5\units{dBm}$. Voltage spacing between the Shapiro steps is calculated as $\Delta V=4.1\units{\mu V}$ (dotted lines). Inset: Enlargement of the $0...5\units{\mu V}$-region shows a weak half-integer Shapiro step. (b) Experimental data of Shapiro step measurements of the same junction at $T=3.2\units{K}$ (underdamped regime). As $f=4\units{GHz}$ a voltage spacing of $\Delta V=8.3\units{\mu V}$ (dotted lines) is expected. Half-integer Shapiro steps are visible and marked by arrows. 
  }
  \label{PRL0_29_1_07_4K_2gigPRL0_24_1_07_300mbar_4gig}
\end{figure}

\begin{figure}[t]
  \includegraphics[width=7.6cm]{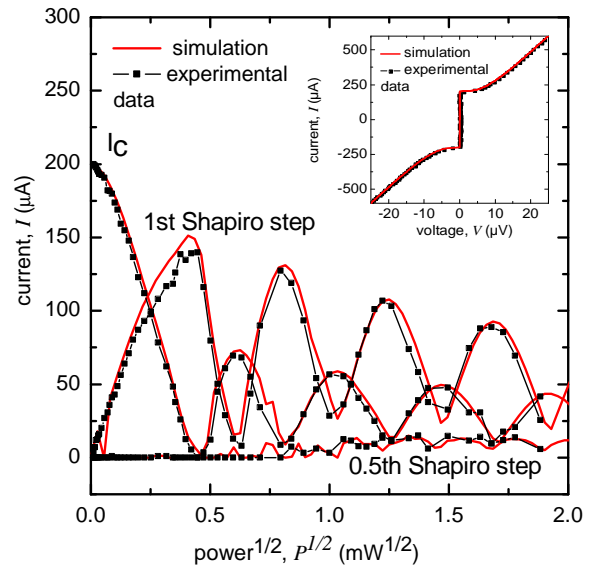}
  \caption{(Color online) 
    Dependence of the height of the 0th (=$I_c$), 1st and 0.5th Shapiro step on the applied microwave power. Experimental data and simulations are compared. Simulations are performed without taking the $\sin(2\phi)$ component into account. Inset: Comparison between experimentally observed IVC and according to the RCSJ-model simulated curve. (sample: $\#1$-$0$ at $T=4.2\units{K}$, $f=2\units{GHz}$, $I_c=202.8\,\mu{\rm A}$, $\beta_c=0.8$)
  }
  \label{PRLnull_29_1_07_auswertung_bessel}
\end{figure}

Shapiro step measurements are carried out for sample set $\#1$. IVCs are measured at different temperatures in the presence of applied microwaves. Constant voltage steps appear on the IVC due to the synchronization of Josephson oscillations to the applied excitation. As an example, Shapiro step measurements for sample $\#1$-$0$ at $T=4.2\units{K}$ and $T=3.2\units{K}$ are shown in Fig.~{\ref{PRL0_29_1_07_4K_2gigPRL0_24_1_07_300mbar_4gig}}. Microwave frequencies of $f=2\units{GHz}$ and $f=4\units{GHz}$ are applied. In both cases resonant steps which fulfill the condition $\Delta V=f \Phi_0$ ($4.1\units{\mu V}$ and $8.3\units{\mu V}$, respectively) are observed. Additionally, half-integer Shapiro steps are visible which have a voltage spacing of $\Delta V=\frac{1}{2}(f \Phi_0$). They are weakly pronounced for $f=2\units{GHz}$ at $T=4.2\units{K}$, see inset of Fig.~{\ref{PRL0_29_1_07_4K_2gigPRL0_24_1_07_300mbar_4gig}} (a). At $T=4.2\units{K}$ the junction is in the overdamped regime. These experimental conditions are chosen to avoid chaotic dynamics. With decreasing temperature and increasing frequency the half-integer Shapiro steps become more pronounced, see Fig.~{\ref{PRL0_29_1_07_4K_2gigPRL0_24_1_07_300mbar_4gig}} (b). 

The occurrence of half-integer Shapiro steps reminds one of the ongoing discussion about non-sinusoidal current-phase relations in \ZeroPi junctions. As already described a $0$ to $\pi$ transition can be achieved in SFS or SIFS junctions as a function of temperature \cite{Ryazanov:2001:SFS-PiJJ} or ferromagnetic barrier thickness.\cite{Kontos:2002:SIFS-PiJJ} As close to the \ZeroPi transition the first order Josephson supercurrent vanishes, the observation of a $\sin(2\phi)$ component seems possible in the transition region. Half-integer Shapiro steps have been reported for SFS junctions in the vicinity of the minimum of $I_c$.\cite{Sellier:2004:SFS:HalfIntShapiro} The authors attributed these steps to the $\sin(2\phi)$ component. In other experiments \cite{Frolov:2006:SFS-0-pi} half-integer Shapiro steps are explained with a non-uniform critical current density, without the need of an intrinsic $\sin(2\phi)$ component. Thus, the conclusive explanation of the origin of half-integer Shapiro steps is still an open problem. 

In order to examine whether our half-integer Shapiro steps are due to a non sinusoidal current-phase relation, Shapiro step measurements for different excitation amplitudes are carried out. As an example, the experimental parameters $I_c$ and $R$ of sample $\#1$-$0$ at $T=4.2\units{K}$ are obtained from experiment and are used for simulations. Current-voltage characteristics can be simulated according to the RCSJ-model, see inset of Fig.~{\ref{PRLnull_29_1_07_auswertung_bessel}}. A current-phase-relation is used with an additional $\sin(2\phi)$ term, which can be weighted by a factor of $\delta$. Shapiro steps are measured in experiment by radiating microwaves with a frequency of $f=2\units{GHz}$. 
As a first try simulations are performed without taking the $\sin(2\phi)$ component into account ($\delta=0$). The height of the Shapiro steps is extracted from experimental data and from simulations. As expected the respective step heights are proportional to the Bessel functions. In Fig.~{\ref{PRLnull_29_1_07_auswertung_bessel}} we compare the height of the Shapiro steps with $n=0,1,\frac{1}{2}$ as a function of applied microwave power, obtained in experiment and simulation. One can see a perfect quantitative agreement. Although the $\sin(2\phi)$ component is not taken into account, the half-integer Shapiro step appears in the simulations and reproduces the experimental data perfectly well. This is a result of a finite capacitance, \ie, $\beta_c\neq 0$.

The simulations are repeated taking the $\sin(2\phi)$ component into account (not shown). As long as the second harmonic is small ($\delta\leq0.1$), the experimental data are reproduced by the numerical simulations. As soon as the second harmonic contribution is higher than $\delta\,\,\textgreater\,\,0.1$, the simulations deviate significantly from the experimental data. These data confirm our assumption, that the half-integer Shapiro steps are not due to a $\sin(2\phi)$ contribution. Half-integer Shapiro steps are also observed for the samples $\#1$-$\pi$ and $\#1$-\ZeroPi (not shown). The experimental data are reproduced in simulations without the need of a $\sin(2\phi)$ component, too. 

Nevertheless, half-integer Shapiro steps are observed here in systems in which the $\sin(2\phi)$ component does not play a dominant role as the $\sin\phi$ component is not suppressed. In this case the half-integer Shapiro step is a subharmonic step and not attributed to a double Josephson frequency.

\section{Conclusions}
\label{Sec:Conclusions}

We experimentally studied the static and \emph{dynamic} properties of $0$, $\pi$ and \ZeroPi SIFS Josephson junctions using two sets of samples: $\#1$ being in the short junction limit, and $\#2$ having an intermediate junction length. All samples can be made underdamped at sufficiently low temperatures below $T=4.2\units{K}$. Fiske steps, zero field steps and Shapiro steps were observed on the $I$-$V$ characteristics. The dynamic properties of 0 and $\pi$ SIFS junctions are qualitatively similar to standard SIS junctions. 

We have observed \emph{half-integer} Shapiro steps on the current-voltage characteristics of $0$, $\pi$ and \ZeroPi Josephson junctions, which does not necessarily imply the presence of the second harmonic in the current-phase relation, but may be present due to the finite capacitance of the Josephson junctions. The analysis of a short overdamped JJ in the framework of the RSJ model, confirms this picture.

\emph{Half-integer} zero field steps were observed on the $I$-$V$ characteristics of \ZeroPi Josephson junctions. The $I_m^{0-\pi}(B_y)$ dependences for several Fiske and zero field steps were measured. To our knowledge, this is the first time when a half-integer zero field step is reported for ``natural'' 0-$\pi$ JJs, as opposed to ``artificial'' 0-$\pi$ JJs with current injectors for which half-integer zero field steps were observed before.\cite{Goldobin:Art-0-pi}

Thus, SIFS $0$, $\pi$ and \ZeroPi JJ technology can already be used to fabricate more complex superconducting electronic devices combining the dynamics of fluxons and semifluxons, \eg, using RSFQ readout for semifluxon bits/qubits or ballistic fluxon readout.\cite{Goldobin:F-SF, last}

\begin{acknowledgments}
  We thank T.~Gaber for useful discussions. Financial support by the the Studienstiftung des Deutschen Volkes (J.~Pfeiffer), by the Evangelisches Studienwerk e.V. Villigst (M.~Kemmler) and by the DFG (project WE 4359/1-1, project SFB/TRR 21 and GO 1106/1-1) is gratefully acknowledged. 
\end{acknowledgments}

\bibliography{this2}

\end{document}